\documentclass[11pt]{article}
\usepackage{amssymb}
\usepackage{amsmath}
\parskip=\medskipamount
\usepackage[dvips]{graphics}

\usepackage{epsfig}

\newtheorem{definition}{Definition}[section]

\newtheorem{theorem}[definition]{Theorem}
\newtheorem{proposition}[definition]{Proposition}

\newtheorem{remark}[definition]{Remark}
\newtheorem{example}[definition]{Example}

\font\fr=eufm10  scaled \magstep 1   
\font\ddpp=msbm10  scaled \magstep 1  
 1  
\def\bull{\ \ \ \vrule height 1.5ex width.8ex depth.3ex \medskip}

\def\QED{\hskip0.1em\hfill\null\ \null\nobreak\hfill
\kern3pt\lower1.8pt\vbox{\hrule\hbox   {\vrule\kern1pt\vbox{\kern1.7pt
\hbox{$\scriptstyle   QED$}\kern0.2pt}\kern1pt\vrule}\hrule}}

\def\R{\hbox{\ddpp R}}               
\def\S{\hbox{\ddpp S}}  

\def\hfl#1#2{\smash{\mathop{\hbox to 12 mm{\rightarrowfill}}
\limits^{\scriptstyle#1}_{\scriptstyle#2}}}

\begin{document}

\title{\bf Geometric integrators and nonholonomic mechanics}

\author{ M. de Le\'on, 
D. Mart{\'\i}n de Diego and A. Santamar{\'\i}a-Merino
\footnote{Instituto de Matem\'aticas y F\'\i sica Fundamental,
CSIC,
Serrano 123, 
28006 Madrid, Spain,
{mdeleon@imaff.cfmac.csic.es}, d.martin@imaff.cfmac.csic.es, aitors@imaff.cfmac.csic.es}
}

%

\maketitle

\begin{abstract}
A geometric derivation of nonholonomic integrators is developed. It is based in the classical technique of generating functions adapted to the special features of nonholonomic systems. The theoretical methodology and   the integrators obtained are  different from the obtained in \cite{JS}. 
In the case of mechanical systems with linear constraints a family of geometric integrators preserving the nonholonomic constraints is given. 

{\sl AMS classification scheme numbers}: 37J60, 58F05, 37M15
\end{abstract}

\section{Introduction}

\subsection{Introduction to nonholonomic mechanics}

The theory of   systems with nonholonomic constrains goes back to the XIX century. D'Alembert's or Lagrange-D'Alembert's  principle of virtual work and Gauss principle of least constraint can be considered to be the first solutions to the analysis of systems with constraints, holonomic or not. After a period of decay, recently many authors show a new interest in that theory and also in its relation  to the new developments in control theory, subriemannian geometry, robotics, etc (see, for instance,\cite{NF}).
The main characteristic of this period  was that Geometry was used in a systematic way (see L.D. Fadeev and A.M. Vershik \ \cite{VF} as an advanced and fundamental reference and, also, \cite{BS,BKMM,CLMD,Cort,KSB,Koiller,KM1,LMD,LD0,LD,Marle1})

As is well known, in most problems of particle mechanics, the motion of the particles is constrained in some way; this is the term used to denote the condition that some motions or configurations are not allowed.
First,  we will start with a configuration space $Q$, which is a $n$-dimensional differentiable manifold, with local coordinates $q^i$.
General two-side or equality constraints are functions of the form 
$\phi^{a}(q^i, \dot{q}^i)=0, 1\leq a\leq m$, 
depending, in general, on configuration coordinates and  their velocities.
The various kinds of constraints we are concerned with will roughly come in two types: holonomic and nonholonomic, depending whether  the constraint  is derived from a constraint in the configuration space or not.
Therefore, the dimension of the space of configurations is reduced by  holonomic constraints   but not by nonholonomic constraints. Thus, holonomic constraints permit a reduction in the number of coordinates of the configuration space needed to formulate a given problem (see \cite{NF}).

We will restrict ourselves to the case  of nonholonomic constraints, since the case of holonomic constraints, and, in particular, the construction of holonomic integrators,  is well established in the existing literature. Geometrically,   nonholonomic constraints are globally described by a submanifold $\tilde{M}$ of the velocity phase space $TQ$, the tangent bundle of the configuration space $Q$.
In case $\tilde{M}$ is a vector subbundle of $TQ$, we are dealing with linear constraints. We will usually refer to $\tilde{M}$ as $D$ and, in such case, the constraints are alternatively defined by a distribution $D$ on the configuration space $Q$. If this distribution is integrable, we are precisely in the case of holonomic constraints. 
In case $\tilde{M}$ is an affine subbundle modeled on a vector bundle $D$, we are in the case of affine constraints. In the sequel, we will denote by $D$ the constraint submanifold on the velocity phase space, no matter if they are determined by linear or nonlinear constraints.

Given the constraints, we need to specify the dynamical evolution of the system. 
The central concepts permitting the extension of mechanics from the Newtonian point of view to the Lagrangian one are the notions of virtual displacements and virtual work; these concepts were formulated in the developments of mechanics, in their application to statics. In nonholonomic dynamics,  the procedure is given by Lagrange-D'Alembert's principle. 
We usually consider nonholonomic constraints of linear  type, which are the constraints that we will regard as natural in a mechanical sense (although the extension for general nonholonomic constraint will be straightforward).
 We now come to the description of the constraint forces; for constraints of that type, Lagrange-D'Alembert's principle allows us to determine the set of possible values of the constraint forces only from the set of admissible kinematic states, that is, from the constraint manifold $D$ determined by the vanishing of the nonholonomic constraints.
Therefore,  assuming that the dynamical properties of the system are mathematically described by a configuration space $Q$,  by a Lagrangian function $L$ and by a distribution determining the linear constraints $D$,  the equations of motion, following Lagrange-D'Alembert's principle,  are 
\begin{equation}\label{b1}
\left[
\frac{d}{dt}\left( \frac{\partial L}{\partial \dot
q^i}\right) - \frac{\partial L}{\partial q^i} \right] 
\delta q^i=0\; ,
\end{equation}
where $\delta q^i$ denotes the virtual displacements verifying
\begin{equation}\label{a111}
\mu^a_i\delta q^i =0 
\end{equation}
and $D^{o}=\hbox{span }\{ \mu^a=\mu^a_idq^i\}$
(for the sake of simplicity, we will assume that the system is not subject to non-conservative forces).
By using the Lagrange multiplier rule we obtain that
\begin{equation}\label{aqq} 
\frac{d}{dt}\left( \frac{\partial L}{\partial \dot
q^i}\right)-\frac{\partial L}{\partial q^i}=\bar{\lambda}_a\mu^a_i \; .
\end{equation}
The term on the right represents the  constraint force or reaction force induced by the constraints. The functions $\bar{\lambda}_a$ are Lagrange multipliers to be determined in order to obtain a set of second order differential equations. These Lagrangian multipliers are computed using  the constraint equations.
An interesting remark,  that  will be used in the sequel,  is  that whenever  the Lagrange multipliers $\bar{\lambda}_a=\bar{\lambda}_a (q^i, \dot q^i)$ have been determined,  then the system of equations (\ref{aqq}) can be considered  a Lagrangian system subject to external conservative forces
given by the right-hand side term, taking, obviously, an initial condition on the constraint submanifold $D$. Automatically, the choice of the Lagrange multipliers $\bar{\lambda}_a$ implies that the solution integral curves also verifies  the constraint equations.

\subsection{Introduction to Geometric Integration and Discrete Mechanics}

Standard methods for simulating the motion of a dynamical system, generically called  numerical integrators, usually take  an initial condition and move it in the direction specified by the equation of motion or an appropriate discretization. But these standard methods ignore all the geometric features of many dynamical systems, as for instance, for Hamiltonian systems we have preservation of the symplectic form, energy (in the autonomous case) and symmetries, if any.
 However, new methods have been recently developed, called geometric integrators, which are concerned with  some of the extra features of  geometric nature of the dynamical systems. Usually, these integrators, in simulations, can  run for long times with lower spurious  effects (for instance, bad energy behavior for conservative systems) than  the traditional ones. As  is well known, the typical test example is the simulation of the solar system. Therefore, there is presently a great interest in geometric integration of differential equations as, for instance, symplectic integrators of Hamiltonian systems \cite{Hair,Sanz}. 

Discrete variational integrators appear as a special kind of geometric integrators. These integrators have  their roots in the optimal control literature  in the 1960's  and 1970's (Jordan and Polack \cite{Jordan}, Cadzow \cite{Cadz}, Maeda \cite{Maed1,Maed2}) and in 1980's by Lee \cite{Lee1,Lee2}, Veselov \cite{Mose,Vese1}.
In these papers, there  appear the discrete action sum, discrete Euler-Lagrange equations,  discrete Noether theorem... Although this kind of symplectic integrators have been considered  for conservative systems 
\cite{Jaro1,Kane1,Mars1,Mars6,Wend1,Wend2},   it  has been recently shown how  discrete variational mechanics can include forced or dissipative systems \cite{Kane3,Mars6}, holonomic constraints \cite{Gonz2,Mars6},  time-dependent systems \cite{LD1,Mars6}, frictional contact \cite{Pand} and nonholonomic constraints (see \cite{Cort,JS}). Moreover, it has been also discussed reduction theory \cite{Bobe,Bobe1,Mars3,Mars4},   extension to field theories \cite{Jaro2,Mars2} and quantum mechanics \cite{Nort}. 
All these integrators have demonstrated exceptionally good longtime behavior and the research of this topic is interesting for numerical and geometric considerations.

Now, we will describe the discrete variational calculus, following the approach in \cite{Wend1} (see also \cite{Baez,Gilli}).
A discrete  Lagrangian 
is a map $L_d:  Q \times Q\rightarrow \R$ (this discrete Lagrangian may be considered as an approximation of the continuous Lagrangian $L: TQ\rightarrow \R$).
Define the action sum $S_d: Q^{N+1}\rightarrow \R$ corresponding to the Lagrangian $L_d$ by
\[
{S_d}=\sum_{k=1}^{N}  L_d(q_{k-1}, q_{k})\; ,
\]
where $q_k\in Q$ for $0\leq k\leq N$. 
For any covector $\alpha\in T_{(x_1,x_2)}^*(Q\times Q)$, we have a decomposition $\alpha=\alpha_1+\alpha_2$ where $\alpha_i\in T^*_{x_i} Q$.
Therefore,
\[
dL_d(q_0, q_1)=D_{1} L_d(q_0, q_1)+D_{2} L_d(q_0,  q_1)\; .
\]
The discrete variational principle or Cadzow's principle  states that the solutions of the discrete system determined by $L_d$ must extremize the action sum given fixed points $q_0$ and $q_N$.
Extremizing ${S_d}$ over $q_k$, $1\leq k\leq N-1$, we obtain the following system of difference equations
\[
 D_1L_d(q_k, q_{k+1})+D_2L_d( q_{k-1}, q_{k})=0\; .
\]
These  equations are usually called the {\em discrete Euler-Lagrange equations}.
Under some regularity hypothesis (the matrix $(D_{12}L_d(q_k, q_{k+1}))$ is regular) this implicit system of difference equations defines a discrete flow
$
\Upsilon:  Q\times Q\longrightarrow Q\times Q$, by 
$\Upsilon(q_{k-1}, q_k)=(q_k, q_{k+1})$. 

The geometrical properties corresponding to this numerical method are obtained defining the discrete Legendre transformation associated to  $L_d$ by
\[
\begin{array}{rrcl}
FL_d:& Q\times Q&\longrightarrow& T^*Q\\
     & (q_0, q_1)&\longmapsto & (q_0, -D_1 L_d(q_0, q_1))\; ,
\end{array}
\]
and the 2-form $\omega_d=FL_d^*\omega_Q$, where $\omega_Q$ is the canonical symplectic form on $T^*Q$. 
The discrete algorithm determined by $\Upsilon$ preserves the symplectic form $\omega_d$, i.e., $\Upsilon^*\omega_d=\omega_d$.
Moreover, if the discrete Lagrangian is invariant under the diagonal action of a Lie group $G$, then the discrete momentum map $J_d: Q\times Q \rightarrow \hbox{\fr g}^*$ defined by
$
\langle J_d(q_k, q_{k+1}), \xi\rangle=\langle D_2L_d(q_k, q_{k+1}), \xi_Q(q_{k+1})\rangle
$
is preserved by the discrete flow. Therefore, these integrators are symplectic-momentum preserving integrators. Here, $\xi_Q$ is the fundamental vector field determined by $\xi\in \hbox{\fr g}$.

Another alternative approach to discrete variational calculus comes from the classical theory of   
generating functions (see, for instance,  \cite{Arno}).  
Since  $(T^*Q, \omega_Q)$ is an exact symplectic manifold, where $\omega_Q$ is the canonical symplectic form of $T^*Q$ and $\omega_Q=-d\theta_Q$, the symplectic flow  $F_h: T^*Q\rightarrow T^*Q$ of a Hamiltonian vector field  $X_H$ is a canonical transformation, and then  $\hbox{Graph}(F_h)$, the graph of $F_h$, is a Lagrangian submanifold  of the symplectic manifold $(T^*Q\times T^*Q, \Omega)$ where ${\Omega}=\pi_2^*\omega_Q-\pi_1^* \omega_Q$. Here, we denote by $\pi_i: T^*Q\times T^*Q\rightarrow T^*Q$, $i=1,2$ the canonical projections. 
Therefore, denoting 
${\Theta}=\pi_2^*\theta_Q-\pi_1^* \theta_Q$ we have that 
\[
i_{F_h}^*{\Omega}=-d i_{F_h}^*\Theta=0\; ,
\]
where $i_{F_h}: \hbox{Graph}(F_h)\mapsto T^*Q\times T^*Q$ is the canonical inclusion. Then, at least locally, there exists a function ${\cal S}^h: \hbox{Graph}(F_h) \rightarrow \R$ such that
$
i_{F_h}^*\Theta=d{\cal S}^h\; .
$
Taking  $(q^i,p_i)$ as natural coordinates in $\hbox{Graph}(F_h)$ and $(q^i, p_i, {\mathbf q}^i, {\mathbf p}_i)$ the coordinates in $T^*Q\times T^*Q$, then, locally ${\cal S}^h$ is a function of $(q,p)$ coordinates. Hence, along $\hbox{Graph}(F_h)$, we have ${\mathbf q}^i={\mathbf q}^i(q,p)$ and ${\mathbf p}^i={\mathbf p}^i(q,p)$ and moreover 
\[
{\mathbf p}_i\,d{\mathbf q}^i-p_idq^i=d{\cal S}^h(q,p)\; . 
\]
Assume that in a neighborhood of some point $x\in \hbox{Graph}(F_h)$, we can change this system of coordinates by new independent coordinates 
$(q^i, {\mathbf q}^i)$ (the local condition is that $\det \left( \partial {\mathbf q}/\partial p\right)\not=0$).
 In such a case, the function ${\cal S}^h$ can be locally expressed as 
$
{\cal S}^h={\cal S}^h(q,p)=S^h(q,{\mathbf q})$.
The function $S^h(q,{\mathbf q})$ will be called a {\em generating function of the first kind} of the canonical transformation $F_h$. Moreover, 
\[
\left\{
\begin{array}{l}
\displaystyle{p_i= -\frac{\partial S^h}{\partial q^i}}\; ,\\
$\,$\\
\displaystyle{{\mathbf p}_i= \frac{\partial S^h}{\partial {\mathbf q}^i}}\; .\\
\end{array}
\right.
\]
A nice and useful interpretation  of the discrete Euler-Lagrange equations is the following theorem  \cite{Kijo,LDA}. 

\begin{theorem}\label{asss}
Let the function $S^{Nh}$ be defined by
\[
S^{Nh}(q_0, q_{N})=\sum_{k=0}^{N-1}S^h(q_k, q_{k+1})\; ,
\]
where $q_k$, $1\leq k\leq N-1$, are stationary points of the right-hand side, that is
\begin{eqnarray}
0&=&D_2 S^h(q_{k-1}, q_k)+D_1 S^h(q_k, q_{k+1})\; , \quad 1\leq k\leq N-1\; ,\label{ooo}  
\end{eqnarray}
then $S^{Nh}$ is a generating function of first class for $F_{Nh}: T^*Q\rightarrow T^*Q$, for $h$ sufficiently small and  where $F_{Nh}$ denotes the flow of $X_H$ over time $Nh$. \end{theorem}

Moreover, if we start with a regular Lagrangian function $L: TQ\rightarrow \R$, and $H: T^*Q\rightarrow \R$ is the locally associated Hamiltonian,  then we also have  the following result (for example, see \cite{LDA})
\begin{proposition}\label{proposition1}
A generating function of the first kind for $F_h$ is given by
\[
S^h(q_0,q_1)=\int^h_0 L(q(t), \dot{q}(t))\, dt\; ,
\]
where $q(t)$ is a solution of the Euler-Lagrange equations such that $q(0)=q_0$ and $q(h)=q_1$.
\end{proposition}

The conclusion is that the discrete variational calculus consists in taking an approximation of the generating function $S^h$. From this approximation  we obtain a new Lagrangian submanifold of $T^*Q\times T^*Q$ and the relation between subsequent steps is given by (\ref{ooo}) for the new generating function,  which are precisely the discrete Euler-Lagrange equations. The symplecticity and preservation of momentum are now direct consequences of this description.

\subsection{Introduction to nonholonomic integrators}

In a recent paper, J. Cort\'es and S. Mart{\'\i}nez \cite{JS} have proposed a  
construction of nonholonomic integrators 
which is  useful for  numerical considerations. Their construction is based on the {\em discrete Lagrange-D'Alembert's principle}. Assuming that the constraints are given by a distribution $D$, this principle states that 
\[
 \left(D_1L_d(q_k, q_{k+1})+D_2L_d( q_{k-1}, q_{k})\right)_i \delta q_k^i=0,\quad 1\leq i\leq N-1\; ,
\]
where $\delta q_k\in D_{q_k}$ and, in addition $(q_k, q_{k+1})\in D_d$. Here $D_d$ denotes a discrete constraint space $D_d\subset Q\times Q$. 
This integrator has a good performance and naturally inherits some geometric properties of the continuous problem. Observe that the method is based on the discretization of the Lagrangian and a coherent discretization of the constraints, and  both determine the discrete constraint forces. 

 Alternatively, we propose a nonholonomic integrator also based on 
the discretization of the Lagrangian function (in a more precise sense, we discretize the action function) but now we take a coherent discretization of the constraint forces  
and  both determine the discrete constraint submanifold. This method gives us, in general, different integrators from those  in \cite{JS}.
The last considerations of the previous section will be our starting point to study nonholonomic integrators, and our equations will be  conceptually equivalent to the proposed for systems with external forces (see \cite{Mars6}).
In the particular case of mechanical systems with linear constraint in the velocities, we study a subclass of our family of nonholonomic integrators with the property of preservation of the original nonholonomic constraints.

\section{Geometrical formulation of nonholonomic systems}

Let $Q$ be a $n$-dimensional differentiable manifold, with local coordinates
$(q^i)$. 
The tangent bundle $TQ$, with induced coordinates $(q^i, \dot q^i)$, is
equipped with two fundamental geometrical objects \cite{LR}: the
Liouville vector field $\Delta$ and the vertical endomorphism $S$. In
natural bundle coordinates we have
\[ 
\Delta = \dot{q}^i \frac{\partial}{\partial \dot{q}^i}, \quad 
S = dq^i \otimes \frac{\partial}{\partial \dot{q}^i}\; .
\]

Consider a Lagrangian system, with Lagrangian $L: TQ \rightarrow \R$, subject
to nonholonomic constraints, defined by a submanifold $D$ of the velocity phase
space $TQ$. We will assume that $\dim D = 2n - m$ and that $D$ is locally 
described by the vanishing of $m$ independent functions $\phi^a$ (the
``constraint functions").

In geometrical terms, D'Alembert's  principle (or Chetaev's principle for nonlinear constraints) implies that the 
constraint forces, regarded as 1-forms on $TQ$ along $D$,  
take their values in the subbundle $S^*(TD^o)$ of $T^*TQ$, where $TD^o$ denotes
the annihilator of $TD$ in $T^*TQ$. 
In an intrinsic way, the equations of motion  can be written as (see \cite{LMD,LD})
\[
\begin{array}{rcl}
(i_X\omega_L-dE_L)_{|D}&\in& S^*(TD^o)\; ,\\
X_{|D} &\in & TD\; ,
\end{array}
\]
where $\omega_L$ is the Poincar\'e-Cartan 2-form defined by
$\omega_L=-d(S^*(dL))$ and $E_L=\Delta(L)-L$ is the energy function.

In the sequel we will also assume that the following {\em admissibility 
condition\/} holds
\[
\dim TD^o=\dim S^*(TD^o)\,.
\] 
This essentially means that the matrix $(\partial \phi^a/\partial \dot{q}^i)$
has rank $m$ everywhere.

We now turn  to the Hamiltonian description of the nonholonomic system on the 
cotangent bundle $T^*Q$ of $Q$ \cite{BS,KM1,Marle1}. The canonical
coordinates on $T^*Q$ are 
denoted by $(q^i, p_i)$, and the cotangent bundle projection will be 
$\pi_Q: T^*Q\rightarrow Q$. Assuming the regularity of the Lagrangian, we have 
that the Lagrangian and Hamiltonian formulations are locally equivalent. If we
suppose, in addition, that the Lagrangian $L$ is hyperregular, then the 
Legendre transformation $Leg: TQ \rightarrow T^*Q, (q^i, \dot{q}^i) \mapsto
(q^i, p_i = \partial L/\partial \dot{q}^i)$, is a global diffeomorphism.
The constraint functions on $T^*Q$ become $\Psi^a=\phi^a\circ Leg^{-1}$, i.e.
\[
\Psi^a(q^i, p_i)=\phi^a(q^i, \frac{\partial H}{\partial p_i})\,,
\]
where the Hamiltonian $H: T^*Q\rightarrow \R$ is defined by  
$H=E_L\circ Leg^{-1}$. 
Since locally 
$Leg^{-1}(q^i, p_i)=(q^i, \displaystyle{\frac{\partial H}{\partial p_i}})$, then
\[
H=p_i \dot{q}^i-L(q^i, \dot{q}^i)\; ,
\]
where $\dot{q}^i$ is expressed in terms of $q^i$ and $p_i$  using $Leg^{-1}$.

The equations of motion for the nonholonomic system on
$T^*Q$ can now be written as follows
\begin{equation}\label{hnh}
\left\{
\begin{array}{rcl}
\dot q^i&=&\displaystyle{\frac{\partial H}{\partial p_i}}\\
\vphantom{\huge A}\dot p_i&=&\displaystyle{-\frac{\partial H}{\partial q^i}-{\lambda}_a \frac{\partial
\Psi^a}{\partial p_j}{\cal H}_{ji}}\, ,
\end{array}\right.
\end{equation}
together with the constraint equations 
$\Psi^a(q,p)=0$, 
where ${\cal H}_{ij}$ are the components of the inverse of the matrix 
$({\cal H}^{ij})=(\partial^2 H/ \partial p_i\partial p_j)$. Note that
\[
(\frac{\partial \Psi^a}{\partial p_j}{\cal H}_{ji})(q,p) = 
(\frac{\partial \phi^a}{\partial \dot{q}^i} \circ Leg^{-1})(q,p).
\]

The symplectic 2-form $\omega_L$ is 
related, via the Legendre map, with the canonical symplectic form $\omega_Q$ on $T^*Q$. Let $M$
denote the image of the constraint submanifold $D$ under the Legendre
transformation, and let $F$ be the distribution 
on $T^*Q$ along $M$, whose annihilator is given by
\[
F^o = Leg_*( S^*(TD^o))\,.
\]
Observe that $F^o$ is locally generated by the $m$ independent 1-forms
\[
\mu^a=\frac{\partial \Psi^a}{\partial p_i}{\cal H}_{ij}d q^j\; ,\ 
1\leq a\leq m \,.
\]
The ``Hamilton equations"  for  the nonholonomic system can be then
rewritten in intrinsic form as
\begin{equation}\label{a1}
\begin{array}{rcl}
(i_X\omega_Q-dH)_{|M}&\in& F^{o}\\
X_{|M} &\in& TM \,.
\end{array}
\end{equation}

Suppose in addition that the following {\em compatibility condition} $F^{\perp}\cap TM=\{0\}$ holds, where $``\perp"$ denotes
the symplectic orthogonal with respect to $\omega_Q$. 
Observe that, locally, this condition means that the matrix
\begin{equation}\label{a2}
({C}^{ab})=
\left(\frac{\partial \Psi^a}{\partial p_i}{\cal H}_{ij}\frac{\partial
\Psi^b}{\partial p_j}\right)
\end{equation}
is regular. 
On the Lagrangian side, the compatibility condition is locally  written as
\begin{equation}\label{a3}
\det (\tilde{C}^{ab})=\det 
\left(\frac{\partial \phi^a}{\partial \dot{q}^i}{W}^{ij}\frac{\partial
\phi^b}{\partial \dot{q}^j}\right)\not=0\; ,
\end{equation}
where $W^{ij}$ are the entries of the Hessian matrix $\displaystyle{\left(\frac{\partial^2 L}{\partial \dot{q}^i\partial\dot{q}^j}\right)_{1\leq i,j\leq n}}$.
The compatibility condition is not too restrictive, since, taking 
into account the admissibility assumption, it is trivially verified by the 
usual systems of mechanical type (i.e.\ with a Lagrangian of the form kinetic
minus potential energy), where the ${\cal H}_{ij}$ represent the components of
a positive definite Riemannian metric. 
The compatibility condition  guarantees in particular the existence of a unique
solution of the constrained  equations of motion (\ref{a1}) which, henceforth,
will be denoted by $X_{H,M}$ on the Hamiltonian side and $\xi_{L,D}$ on the Lagrangian side.

Moreover, if we denote by $X_H$ the Hamiltonian vector field of $H$, i.e.,
$i_{X_H}\omega_Q=dH$ then, using the constraint functions, we may explicitely determine the
Lagrange multipliers $\lambda_a$ as
\[
\lambda_a=  -{\cal C}_{ab} X_H(\Psi^b)\; .
\]
Next, writing the 1-form  
\[
\Lambda=-{\cal C}_{ab} X_H(\Psi^b)\frac{\partial
\Psi^a}{\partial p_j}{\cal H}_{ji}dq^i\; ,
\]
the nonholonomic equations are equivalently rewritten as
\begin{equation}\label{hnh1}
\left\{
\begin{array}{rcl}
\dot q^i&=&\displaystyle{\frac{\partial H}{\partial p_i}}\; ,\\
\dot p_i&=&\displaystyle{-\frac{\partial H}{\partial q^i}-\Lambda_i}\, ,
\end{array}\right.
\end{equation}
for initial conditions $(q_0, p_0)\in M$ and $\Lambda=\Lambda_i\, dq^i$.
We also denote by $\tilde{\Lambda}={Leg}^*({\Lambda})$ the 1-form on $TQ$ wich represents the constraint force   
once the Lagrange multipliers have been determined.

Now, consider the flow 
$F_t: M\rightarrow M$, $t\in I\subseteq \R$  of the vector field $X_{H, M}$, solution of the nonholonomic problem.

Since (\ref{hnh1}) is geometrically rewritten as
\[
i_{X_{H, M}}\omega_Q=dH+\Lambda\; ,
\]
($i_{\xi_{L,D}}\omega_L=dE_L+\tilde{\Lambda}$, with $\tilde{\Lambda}=Leg^*\Lambda$, on the Lagrangian side)
then
\[
L_{X_{H, M}}\theta_Q=d(i_{X_{H,M}}\theta_Q-H)-\Lambda\; ,
\]
or, equivalently,
\[
L_{X_{H, M}}\theta_Q=d(L\circ Leg^{-1})-\Lambda\; .
\]
Now, from the dynamical definition of the Lie derivative, we have
\[
F_t^*\left(L_{X_{H, M}}\theta_Q\right)=\frac{d}{dt}\left( F_t^*\theta_Q\right)\, ,
\]
and integrating, we obtain  the following expression, with some abuse of notation,

\begin{equation}\label{ei}
F_h^*\theta_Q-\theta_Q=d\left(\int^h_0 L\circ \tilde{F}_t \,dt\right)-\int^h_0 F_t^*\Lambda\; ,
\end{equation}
where $\tilde{F}_t$ is the flow of the vector field $\xi_{L, D}$.
In next sections, we will study geometric integrators which verify a discrete version of equation (\ref{ei}).

\section{``Generating  functions" and nonholonomic mechanics}

Next, we will follow similar arguments for the construction of generating functions for symplectic or canonical maps \cite{Arno}. However, because of equation (\ref{ei}), we have that the nonholonomic flow is not a canonical transformation; i.e.,
\begin{equation}\label{ei1}
F_h^*\omega_Q-\omega_Q=d\left(\int^h_0 F_t^*\Lambda\right)\; .
\end{equation}
 This description will allow us to construct a new family of nonholonomic integrators for equations (\ref{aqq}). 
Denote by $\pi_i: T^*Q\times T^*Q\rightarrow T^*Q$, $i=1,2$, the canonical projections. Consider the following forms
\begin{eqnarray*}
{\Theta}&=&\pi_2^*\theta_Q-\pi_1^* \theta_Q\; ,\\
{\Omega}&=&\pi_2^*\omega_Q-\pi_1^* \omega_Q=-d{\Theta}\; .\\
\end{eqnarray*}
Denote  by $i_{F_h}: \hbox{Graph}(F_h)\hookrightarrow T^*Q\times T^*Q$ the inclusion map and observe  that
$\hbox{Graph}(F_h)\subset M\times M$.
Then, from (\ref{ei1})   
\begin{eqnarray*}
i_{F_h}^*{\Omega}&=&({\pi_1}_{|\hbox{\footnotesize Graph}(F_h)})^*(F_h^*\omega_Q-\omega_Q)\\
&=& ({\pi_1}_{|\hbox{\footnotesize Graph}(F_h)})^*\left[d\left(\int^h_0 F_t^*\Lambda\right)\right]\; ,
\end{eqnarray*}
or, from (\ref{ei}),
\[
i_{F_h}^*{\Theta}=({\pi_1}_{|\hbox{\footnotesize  Graph}(F_h)})^*\left[ d\left(\int^h_0 L\circ \tilde{F}_t \,dt\right)-\int^h_0 F_t^*{\Lambda}\right]\; .
\]


Let  $(q_0, p_0, q_1, p_1)$  be coordinates in $T^*Q\times T^*Q$ in a neighborhood of some point in  $\hbox{Graph}(F_h)$. If   $(q_0, p_0, q_1, p_1)\in \hbox{Graph}(F_h)$ then  
$\Psi^a(q_0, p_0)=0$ and $\Psi^a(q_1,  p_1)=0$.
Moreover,  along $\hbox{Graph}(F_h)$, $q_1=q_1(q_0,p_0)$ and $p_1=p_1(q_0,p_0)$, 
\begin{equation}\label{fr}
p_1\,dq_1-p_0dq_0=d \left(\int^h_0 L(q(t), \dot{q}(t)) \,dt\right)-\int^h_0 \widetilde{\Lambda}(q(t),\dot{q}(t))\; , 
\end{equation}
where $(q(t), \dot{q}(t))=\tilde{F}_t(q_0, \dot{q}_0)$ with $Leg(q_0, \dot{q}_0)=(q_0, p_0)$. Here, $\tilde{F}_t$ denotes the flow of $\xi_{L,D}$. Equation (\ref{fr}) is satisfied along $\hbox{Graph}(F_h)$.

Assume that, in a neighborhood of some point $x\in \hbox{Graph}(F_h)$, we can change this system of coordinates to a  new  coordinates 
$(q_0, q_1)$. 
Denote by 
\[
S^h(q_0, q_1)=\int^h_0 L(q(t), \dot{q}(t))\, dt\, ,
\]
where $q(t)$ is a solution curve  of the nonholonomic problem with $q(0)=q$ and $q(h)=q_1$. This solution always exists for adequate values of $q_0$ and $q_1$. In fact, observe that
\[
q_1=q_0+h\frac{\partial H}{\partial p}(q_0, p_0)+o(h^2)\; ,
\]
hence, since 
  $\det \left(\frac{\partial^2 H}{\partial p_i\partial p_j}\right)\not=0$, we locally have that  
$p_0=p_0(q_0, q_1,h)$. But, in addition, $(q_0, p_0)\in M$; therefore $\varphi^a(q_0, q_1, h)=\Psi^a(q_0, p_0(q_0, q_1, h))=0$.
Then, the curve 
$$(q(t), \dot{q}(t))=Leg^{-1}({F}_t(q_0, p_0(q_0, q_1,h)))\; ,$$ 
verifies the required assumptions if $\varphi^a(q_0, q_1, h)=0$. 

Thus,  we deduce  
 that\footnote{For a function $f({x}, {y})$ with ${x}, {y}\in \R^n$ we use the notation ${\partial f}/{\partial {x}}$ (respectively, ${\partial f}/{\partial {y}}$) to write the partial derivative with respect the first $n$-variables (resp., the second $n$-variables).}
\begin{equation}\label{aq1}
\left\{
\begin{array}{l}
\displaystyle{p_0= -\frac{\partial S^h}{\partial q_0}}+\int^h_0 \widetilde{\Lambda}(q(t),\dot{q}(t))\frac{\partial q}{\partial q_0}\; ,\\
$\,$ \\
\displaystyle{p_1= \frac{\partial S^h}{\partial q_1}}-\int^h_0 \widetilde{\Lambda}(q(t),\dot{q}(t))\frac{\partial q}{\partial q_1}\; ,\\
\end{array}
\right.
\end{equation}
where $(q_0, q_1)$  verifies the constraint functions $\varphi^a(q_0, q_1,h)=0$, now explicitely  defined by 
\begin{equation}\label{qwe}
\hskip-0.5cm\varphi^a(q_0, q_1,h)= \Psi^a(q_0, -\frac{\partial S^h}{\partial q_0}(q_0, q_1)+\int^h_0 \widetilde{\Lambda}(q(t),\dot{q}(t))\frac{\partial q}{\partial q_0})\; ,\quad 1\leq a\leq m\; ,
\end{equation}
with $q(t)$ solution of the nonholonomic problem with $q(0)=q_0$ and $q(h)=q_h$.

Next, we will show how the group composite law of the flow $F_h$
\[
F_{Nh}=\underbrace{F_h\circ\ldots \circ F_h}_{N}
\]
is expressed in terms of the corresponding ``generating functions" $S^h$. 
Moreover, the following Theorem will result in  a new construction of numerical integrators for nonholonomic mechanics when we change the ``generating function" and the constraint forces by appropriate approximations. 
As a generalization of Theorem \ref{asss} we have the following 

\begin{theorem}\label{Th}
The function $S^{Nh}$, the ``generating function" for $F_{Nh}$, is  given  by
\[
S^{Nh}(q_0, q_{N})=\sum_{k=0}^{N-1}S^h(q_k, q_{k+1})\; ,
\]
where $q_k$, $1\leq k\leq N-1$, are  points verifying
\begin{equation}\label{ase}
\hskip-2cm D_2 S^h(q_{k-1}, q_k)+D_1 S^h(q_k, q_{k+1})=
\int^h_0\widetilde{\Lambda}(q(t), \dot{q}(t))\frac{\partial q}{\partial q_1}+\int^{2h}_h\widetilde{\Lambda}(q(t), \dot{q}(t))\frac{\partial q}{\partial q_0}\; ,
\end{equation}
and
$q(t)$ is a solution curve  of the nonholonomic problem with $q(0)=q_{k-1}$ and $q(h)=q_{k}$ 
(respectively, $q(h)=q_k$ and $q(2h)=q_{k+1}$) for the first integral (resp., second integral) of the right-hand side.
\end{theorem}
{\bf Proof:}
It is suffices to prove the result for $N=2$; that is, 
\[
S^{2h}(q_0, q_2)=S^h(q_0, q_1)+S^h(q_1, q_2)\; ,
\]
where $q_1$ verifies condition (\ref{ase}).

Since
\begin{eqnarray*}
p_1\,dq_1-p_0\, dq_0&=&dS^h(q_0, q_1)-\int^h_0 \widetilde{\Lambda}(q(t),\dot{q}(t))\; ,\\
p_{2}\,dq_{2}-p_1\, dq_1&=&dS^h(q_1, q_2)-\int^{2h}_h \widetilde{\Lambda}(q(t),\dot{q}(t))\; ,
\end{eqnarray*}
then 
\[
\hskip-1.5cm p_2\,dq_2-p_0\, dq_0=d\left(S^h(q_0, q_1)+S^h(q_1, q_2)\right)-\int^h_0 \widetilde{\Lambda}(q(t),\dot{q}(t))-\int^{2h}_h \widetilde{\Lambda}(q(t),\dot{q}(t))\; .
\]
Since the variables $q_1$ do not appear on the left-hand side term, it follows that
\begin{equation}\label{impo}
\hskip-2cm 0=D_2 S^h_1(q_{0}, q_1)+D_1 S^h_2(q_1, q_{2})-\int^h_0 \widetilde{\Lambda}(q(t),\dot{q}(t))\frac{\partial q}{\partial q_1}
-\int^{2h}_{h} \widetilde{\Lambda}(q(t),\dot{q}(t))\frac{\partial q}{\partial q_0}\; ,
\end{equation}
and for a choice of $q_1$ verifying (\ref{impo})  then 
\[
S^{2h}(q_0, q_2)=S^h(q_0, q_1)+S^h(q_1, q_2)
\]
is a ``generating function of the first kind" of $F_{2h}$ because
\[
p_2\,dq_2-p_0\, dq_0=dS^{2h}(q_0, q_2)-\int^{2h}_0 \widetilde{\Lambda}(q(t),\dot{q}(t))\; .
\hfill \bull
\]

Equations (\ref{ase}) determine a implicit system of difference equations which permit us to obtain $q_2$ from the initial data $q_0$ and $q_1$. An interesting consequence
 is that these equations preserve the constraint submanifold determined by the constraints $\varphi^a=0$, $1\leq a\leq m$. In fact, if $\varphi^a(q_0, q_1,h)=0$ (that is $\Psi^a(q_0, p_0)=0$) then 
\[
\varphi^a(q_1, q_2,h)=\Psi^a(q_1,   \frac{\partial S^h}{\partial q_1}(q_0, q_1)-\int^{h}_{0} \widetilde{\Lambda}(q(t),\dot{q}(t))\frac{\partial q}{\partial q_1})\; ,
\]
and now applying (\ref{aq1}) we obtain that  
\[
\varphi^a(q_1, q_2, h)= \Psi^a(q_1, p_1)=0\; ,
\]
since $F_h(q_0, p_0)=(q_1,p_1)$ and the flow preserves the constraints.

The next remark will be a key result for the construction of nonholonomic integrators. 
\begin{remark}\label{wer}
{\rm
Replace equation (\ref{aq1}) by 
\begin{equation}\label{aq2}
\left\{
\begin{array}{l}
\displaystyle{p_0= -\frac{\partial \tilde{S}^h}{\partial q_0}}+\alpha^h_0(q_0, q_1)\; ,
$\,$\\
\displaystyle{p_1= \frac{\partial \tilde{S}^h}{\partial q_1}}-\alpha^h_1(q_0, q_1)\; ,
\end{array}
\right.
\end{equation}
where $\tilde{S}^h$ is a function of $(q_0, q_1)$ coordinates and $\alpha^h=\alpha^h_0\, dq_0+\alpha^h_1\,dq_1$
and replace the constraints functions 
by 
\begin{equation}\label{qwe1}
{\tilde{\varphi}}^a(q_0, q_1,h)= \Psi^a(q_0, -\frac{\partial \tilde{S}^h}{\partial q^0}+\alpha^h_0(q_0, q_1))\; ,
\end{equation}
that is, 
\[
p_1\, dq_1-p_0\,dq_0=d\tilde{S}^h-\alpha^h\; ,
\]
along $\tilde{\varphi}^a=0$.

Assume that  
\begin{equation}\label{poi}
\det \left(\frac{\partial^2 \tilde{S^h}}{\partial q_0\partial q_1}-\frac{\partial\alpha^h_0}{\partial q_1}\right)\not=0\; ,
\end{equation}
then, applying the implicit function theorem we have that, locally, $q_1=q_1(q_0, p_0)$, and then  the  mapping  
\[
G_h(q_0, p_0)=(q_1, p_1)
\]
is well-defined.

Consider the mapping $G_{Nh}$ defined by 
\[
G_{Nh}=\underbrace{G_h\circ\ldots \circ G_h}_{N}\; .
\]
Following a similar argument to Theorem \ref{Th},  $\hbox{Graph}(G_{Nh})$ is described by 
\begin{equation}\label{aq10}
\left\{
\begin{array}{l}
\displaystyle{p_0= -\frac{\partial \tilde{S}^{Nh}}{\partial q_0}}(q_0, q_N)+\alpha^{Nh}_0(q_0, q_N)\; ,
$\,$\\
\displaystyle{p_N= \frac{\partial \tilde{S}^{Nh}}{\partial q_N}}(q_0, q_N)-\alpha^{Nh}_1(q_0, q_N)\; ,
\end{array}
\right.
\end{equation}
where
$\tilde{S}^{Nh}(q_0, q_{N})=\sum_{k=0}^{N-1}\tilde{S}^h(q_k, q_{k+1})$ and 
$\alpha^{Nh}(q_0, q_{N})=\sum_{k=0}^{N-1}\alpha^h(q_k, q_{k+1})$. Here,  the $q_k$'s, $1\leq k\leq N-1$, verify
 \begin{equation}\label{ase2}
\hskip-2cm D_2 \tilde{S}^h(q_{k-1}, q_k)+D_1 \tilde{S}^h(q_k, q_{k+1})=
\alpha^h_1(q_{k-1}, q_k)+\alpha^h_0(q_k, q_{k+1}), \quad 1\leq k\leq N-1\; . 
\end{equation}

}
\end{remark}

\subsection{Constraint error analysis}

As we have seen, if our ``generating function" is $S^{h}$, then we have exact preservation of the constraints $\varphi^a$. We now investigate what happens when the ``generating function"  is  an approximation. We follow similar arguments to those in  subsection 2.3.1 in \cite{Mars6}.

Assume that $Q$, and also $TQ$ and $T^*Q$, are  finite-dimensional vector spaces with inner product $\langle .,.\rangle$ and corresponding norm $\|\;\|$.  

Consider an ``approximated generating function" $\tilde{S}^h$ and an approximated discrete constraint force $\alpha^h=\alpha_i^h\, dq^i$ for the nonholonomic problem both of order $r$; hence,  there exists an open set $U\subset D$ with compact closure and  constants $c, d_i>0$, $1\leq i\leq n$,  and $H>0$ such that
\begin{eqnarray}
\tilde{S}^h(q_0, q_1)&=&{S}^h(q_0, q_1) +C(q_0, q_1,h)  h^{r+1}\label{rt1}\\
\alpha_i^h&=&\int^h_0 \tilde{\Lambda}_i(q(t), \dot{q}(t))\, dt+ D_i(q_0, q_1, h) h^{r+1}\label{rt2}
\end{eqnarray}
for all solution $q(t)$ of the nonholonomic problem with $q(0)=q_0$, $q(h)=q_1$ and initial condition belonging to $U$ and $h\leq H$. Here 
$C$ and $D_i$, $1\leq i\leq n$, are functions such that $\| C(q_0, q_1, h)\|\leq c$ and 
$\| D_i(q_0, q_1, h)\|\leq d_i$ on $U$.

Taking derivatives we have that 
 \[
\frac{\partial \tilde{S}^h}{\partial q_0}(q_0, q_1)=\frac{\partial\tilde{S}^h}{\partial q_0}(q_0, q_1) +\frac{\partial C}{\partial q_0}(q_0, q_1, h)  h^{r+1}
\]
and also  
\[
\alpha^h_0(q_0, q_1)=(\alpha_0)_i^h\frac{\partial q^i}{\partial q_0}=\int^h_0 \tilde{\Lambda}_i(q(t), \dot{q}(t))\frac{\partial q^i}{\partial q_0}\, dt+\sum_{i=1}^n\frac{\partial D_i}{\partial q_0}(q_0, q_1, h) h^{r+1}
\]
where now $\alpha^h=\alpha^h_0\,  dq_0+\alpha^h_1\, dq_1$

Therefore, we deduce that 
\begin{eqnarray*}
\tilde{\varphi}^a(q_0, q_1,h)&=&\Psi^a(q_0,  -\frac{\partial \tilde{S}}{\partial q^0}+\alpha_0(q_0, q_1))\\
&=& \Psi^a(q_0, -\frac{\partial S^h}{\partial q_0}+\int^h_0 \widetilde{\Lambda}(q(t),\dot{q}(t))\frac{\partial q}{\partial q_0})+E^a(q_0, q_1, h)h^{r+1}\\
&=&\Psi^a(q_0, p_0)+E^a(q_0, q_1, h)h^{r+1}=E^a(q_0, q_1, h)h^{r+1}
\end{eqnarray*}
where $E^a$ are bounded functions. 
Then, the discrete algorithm   preserves the constraints up to order $r$.

\subsection{Local error analysis}
Assuming that 
\begin{equation}\label{poi1}
\det \left(\frac{\partial^2 \tilde{S}^h}{\partial q_0\partial q_1}-\frac{\partial\alpha^h_0}{\partial q_1}\right)\not=0\; ,
\end{equation}
we obtain a discrete flow $G^h: V\subseteq M\longrightarrow M$.
It is easy to show,  from conditions (\ref{rt1}) and (\ref{rt2}), that $G_h$ is an integrator of $X_{H, M}$ of order $r$, following similar arguments to those used in the  subsection above (see also Theorem 2.3.1., in \cite{Mars6}).

\section{ Nonholonomic integrators}\label{nh}

In the sequel and  for simplicity assume that $Q$ is a vector space.
Since $S^h(q_0, q_1)=\int^h_0 L(q(t), \dot{q}(t))\, dt$, where $q(t)$ is a nonholonomic solution with $q(0)=q_0$ and $q(h)=q_1$,  using Remark \ref{wer}, 
we can obtain nonholonomic integrators by taking adequate approximations of the ``generating function" $S^h$ and the extra-term 
$\int^h_0\tilde{\Lambda}(q(t), \dot{q}(t))$.

Consider, for instance,  the approximation
\begin{equation}\label{lol}
S^h_{\alpha}(q_0, q_1)=hL((1-\alpha)q_0+\alpha q_1, \frac{q_1-q_0}{h})\; ,
\end{equation}
for some parameter $\alpha\in [0,1]$.  (In general, we will write $S^h_{\alpha}(q_0, q_1)\approx S^h(q_0, q_1)$.)

A natural approximation of the constraint forces adapted to our choice of approximation for $S^h$ are
\begin{eqnarray*}
\int^h_0\widetilde{\Lambda}(q(t), \dot{q}(t))\frac{\partial q}{\partial q_0}&\approx& 
(1-\alpha) h \widetilde{\Lambda}((1-\alpha)q_0+\alpha q_1, \frac{q_1-q_0}{h})\; ,\\
\int^{h}_{0}\widetilde{\Lambda}(q(t), \dot{q}(t))\frac{\partial q}{\partial q_1}&\approx& 
\alpha h \widetilde{\Lambda}((1-\alpha)q_0+\alpha q_1, \frac{q_1-q_0}{h})\; .\\
\end{eqnarray*}

Consequently, equations (\ref{ase2}) give us the following numerical method for nonholonomic systems 

\fbox{\parbox{16cm}{
\begin{eqnarray*}\label{ase1}
&&D_2 S^h_{\alpha}(q_{k-1}, q_k)+D_1 S^h_{\alpha}(q_k, q_{k+1})=
\alpha h \widetilde{\Lambda}((1-\alpha)q_{k-1}+\alpha q_k, \frac{q_k-q_{k-1}}{h})\\
&&\qquad\qquad +(1-\alpha) h \widetilde{\Lambda}((1-\alpha)q_k+\alpha q_{k+1}, \frac{q_{k+1}-q_k}{h})\; ,\quad 1\leq k\leq N-1\; ,
\end{eqnarray*}
}}
with initial condition satisfying 
\[
\tilde{\varphi}^a(q_0, q_1, h)=\Psi^a(q_0, -\frac{\partial S^h_{\alpha}}{\partial q_0}(q_0, q_1)+ 
(1-\alpha) h \widetilde{\Lambda}((1-\alpha)q_0+\alpha q_1, \frac{q_1-q_0}{h}))=0\; .
\]

\begin{remark}
{\rm 
Obviously, it is possible to produce a wider variety of discrete methods. For example, 
\[
S^h_{\hbox{\footnotesize sym}, \alpha}=\frac{1}{2}S^h_{\alpha}+\frac{1}{2}S^h_{1-\alpha}\; ,
\]
gives a second-order method for any $\alpha\in [0,1]$. Also, higher-order approximations of the function $S^h$ may be considered. 
}
\end{remark}

\begin{example}
{\rm {\bf Nonholonomic particle}.

Consider the Lagrangian $L: T\R^3\rightarrow \R$
\[
L=\frac{1}{2}(\dot{x}^2+\dot{y}^2+\dot{z}^2)-(x^2+y^2)\; ,
\]
 subject to the constraint
\[
\phi=\dot{z}-y\dot{x}=0\; .
\]
It is easy to compute the nonholonomic differential equations 
\begin{eqnarray*}
\ddot{x}&=&-\frac{2x+y\dot{x}\dot{y}}{1+y^2}\\
\ddot{y}&=&-2y\\
\ddot{z}&=&\frac{-2xy+\dot{x}\dot{y}}{1+y^2}\; ,
\end{eqnarray*}
where now  the constraint 1-form is 
\[
\tilde{\Lambda}=\frac{2xy-\dot{x}\dot{y}}{1+y^2}(dz-ydx)\; .
\]

Taking 
\begin{eqnarray*}
S^h_{1/2}(x_0, y_0, z_0, x_1, y_1, z_1)&=&\frac{h}{2}\left[\left(\frac{x_1-x_0}{h}\right)^2+
\left(\frac{y_1-y_0}{h}\right)^2+\left(\frac{z_1-z_0}{h}\right)^2\right]\\&&
-\left(\frac{x_0+x_1}{2}\right)^2-\left(\frac{y_0+y_1}{2}\right)^2\; ,
\end{eqnarray*}
we obtain
 the nonholonomic integrator
\begin{eqnarray*}
&&\hskip-2.5cm\frac{x_1-x_0}{h}-h\frac{x_1+x_0}{2}-\frac{x_2-x_1}{h}-h\frac{x_2+x_1}{2}\\
&&\hskip-2.3cm= -\frac{h}{2}\left[
\frac{
\frac{(x_1+x_0)(y_1+y_0)}{2}
-\frac{(x_1-x_0)(y_1-y_0)}{h^2}}
{1+\left(\frac{y_1+y_0}{2}\right)^2}\cdot \frac{y_1+y_0}{2}
+\frac{
\frac{(x_2+x_1)(y_2+y_1)}{2}-\frac{(x_2-x_1)(y_2-y_1)}{h^2}\; ,
}
{1+\left(\frac{y_2+y_1}{2}\right)^2}\cdot \frac{y_2+y_1}{2}\right]\\
&&\hskip-2.5cm\frac{y_1-y_0}{h}-h\frac{y_1+y_0}{2}-\frac{y_2-y_1}{h}-h\frac{y_2+y_1}{2}=0\; ,\\
&&\hskip-2.5cm\frac{z_1-z_0}{h}-\frac{z_2-z_1}{h}\\
&&\hskip-2.3cm= \frac{h}{2}\left[
\frac{
\frac{(x_1+x_0)(y_1+y_0)}{2}
-\frac{(x_1-x_0)(y_1-y_0)}{h^2}}
{1+\left(\frac{y_1+y_0}{2}\right)^2}
+\frac{
\frac{(x_2+x_1)(y_2+y_1)}{2}-\frac{(x_2-x_1)(y_2-y_1)}{h^2}
}
{1+\left(\frac{y_2+y_1}{2}\right)^2}\right]\; .\\
\end{eqnarray*}
The constraint function on $\R^3\times \R^3$ is  
\begin{eqnarray*}
&&\hskip-1.5cm\tilde{\varphi}^a(x_0, y_0, z_0, x_1, y_1, z_1, h)=
-\frac{z_1-z_0}{h}-\frac{h}{2}
\frac{
\frac{(x_1+x_0)(y_1+y_0)}{2}
-\frac{(x_1-x_0)(y_1-y_0)}{h^2}}
{1+\left(\frac{y_1+y_0}{2}\right)^2}
\\&+&y_0\left[
\frac{x_1-x_0}{h}+h\frac{x_1+x_0}{2}
-
\frac{h}{2}
\frac{
\frac{(x_1+x_0)(y_1+y_0)}{2}
-\frac{(x_1-x_0)(y_1-y_0)}{h^2}}
{1+\left(\frac{y_1+y_0}{2}\right)^2}\cdot \frac{y_1+y_0}{2}
\right]\; .
\end{eqnarray*}

The following two figures show the preservation of energy as a key
point of comparison of computational implementations of the method
exposed above to other methods.

The first figure compares the method introduced here to the
traditional Runge-Kutta method of fourth order, showing an
improvement in several orders of magnitude. Observe that, in this scale,  the value of the energy in each step of our algorithm is practically undistinguishable from the initial value of the energy.
\begin{center}
\includegraphics[width=10cm]{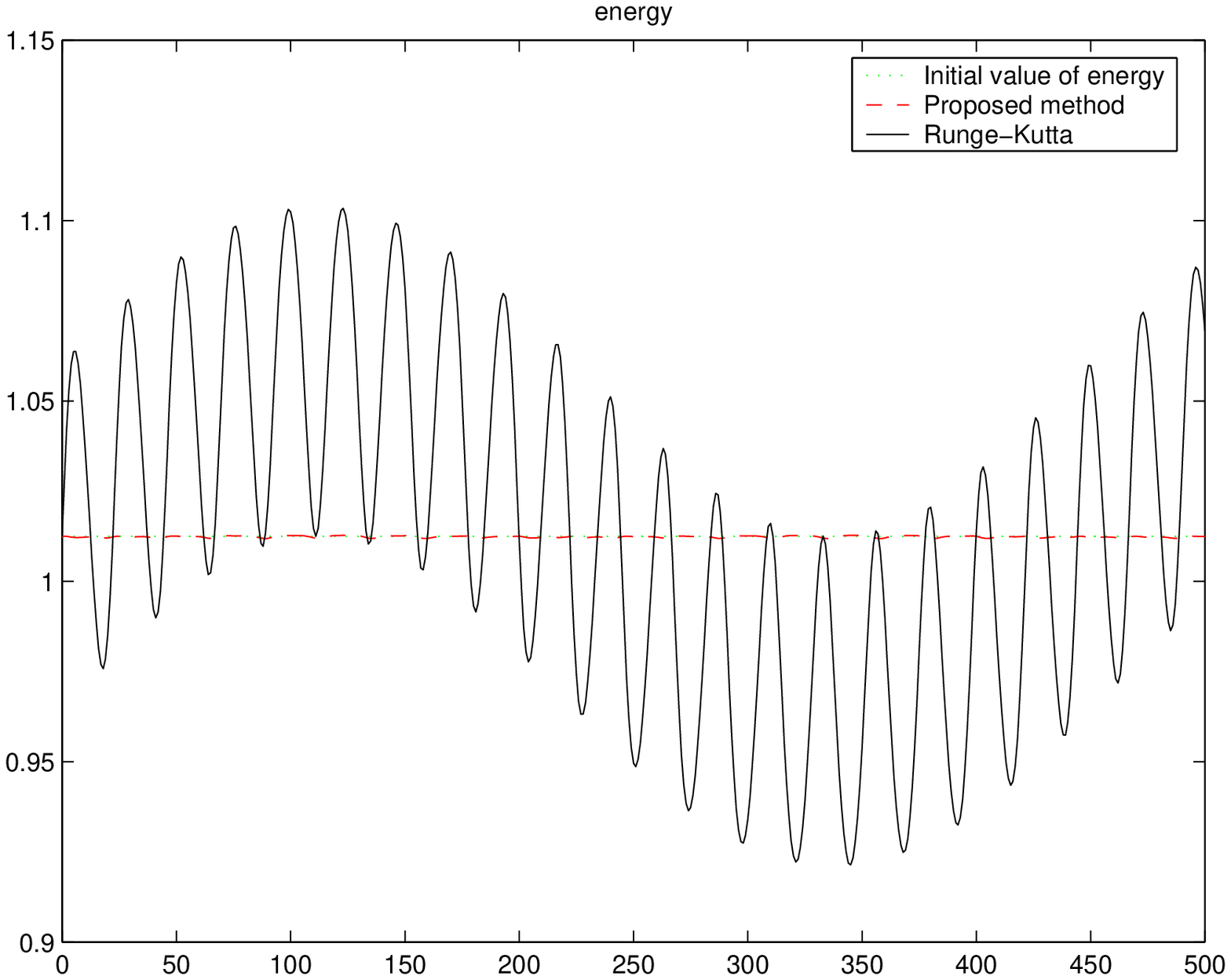}
\end{center}

The second figure is a comparison between our method and the one
appeared in \cite{Cort,JS}. A similar behaviour is observed. 
Nevertheless, a slightly better behaviour can also be appreciated,
where the proposed algorithm shows on average a better
preservation of the original energy.

\begin{center}
\includegraphics[width=10cm]{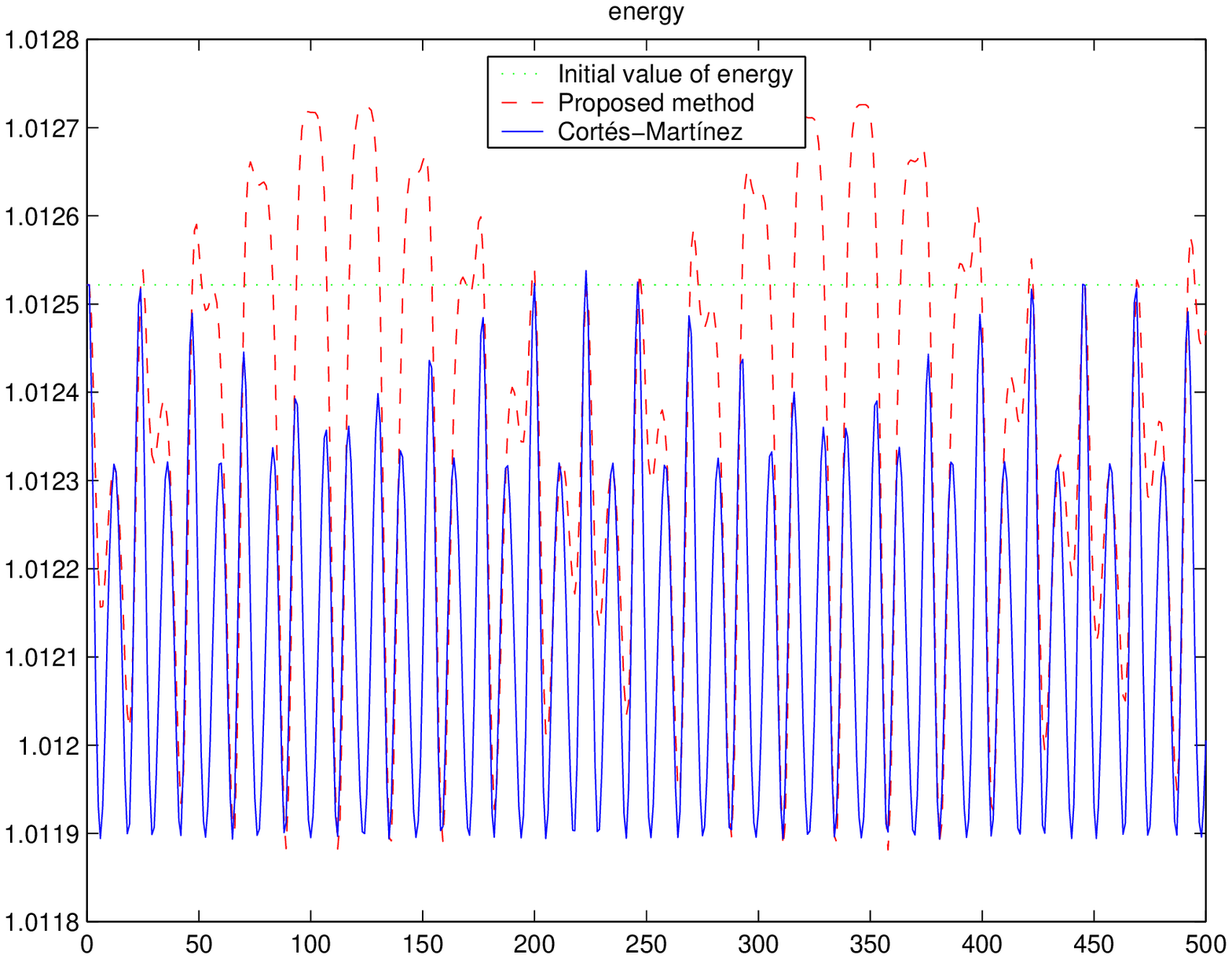}
\end{center}

For the same initial conditions and data, the following graph shows a 
very good behaviour of the constraint function evolution with time 
(notice the small scale).

\begin{center}
\includegraphics[width=10cm]{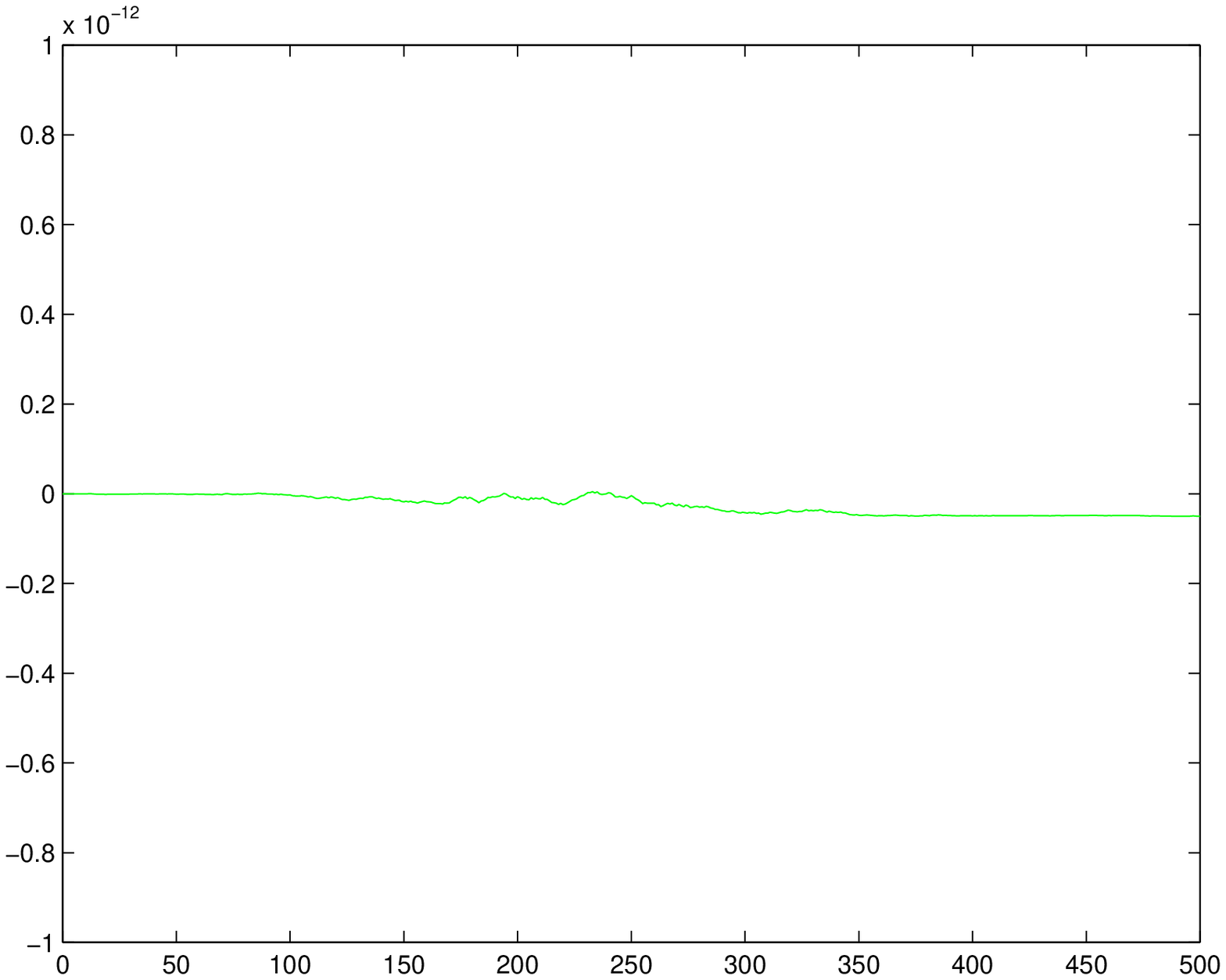}
\end{center}

}
\end{example}

\section{Mechanical systems with linear constraints. Geometric numerical methods preserving constraints}

Suppose  that the mechanical system, given by the Lagrangian 
$L: TQ\rightarrow \R$
\[
L(v_q)=\frac{1}{2}g(v_q, v_q)-V(q)
\]
 is subjected to nonholonomic
constraints $\phi^{a} : TQ \longrightarrow \R$, $1\leq a\leq m$.
Since the nonholonomic constraints usually found in mechanics are linear in 
the velocities we will assume that
\[
\phi^{a} (q,\dot q)= \mu_i^{a} (q) \dot{q}^i , \ 1\leq a \leq m\; .
\]
From a  geometric point of view, these linear constraints are
determined by prescribing a distribution ${\cal D}$ on $Q$ of dimension
$n-m$ such that the annihilator of ${\cal D}$ is locally given by 
\[
{\cal D}^o = \langle \mu^{a}=\mu_i^{a}dq^i \; ; 1 \leq a \leq m \rangle \, .
\]
In this manner, the solutions of the nonholonomic Lagrangian system
satisfy  
\begin{equation}\label{mnb}
\nabla _{\dot{c}(t)} \dot{c}(t) = - {\rm grad}~
V(c(t))+\lambda(\dot{c}(t)),\quad \dot{c}(t) \in {\cal D}_{c(t)}\; ,
\end{equation}
where $\lambda$ is a section of ${\cal D}^{\perp}$ along $c$, and 
${\cal D}^{\perp}$ stands for the orthogonal complement of ${\cal D}$
with respect to the metric $g$.

Since $g$ is a Riemannian metric, the $m \times m$ matrix
$(C^{ab})=(\mu^a_ig^{ij}\mu^b_j)$ is symmetric and regular.
Therefore, we can explicitly determine 
\begin{equation}\label{qqqq}
\lambda(q^{i}(t),\dot{q}^{i}(t)) =
C_{ab} \left((-\Gamma^i_{jk}\dot{q}^j\dot{q}^k - g^{ij} \frac{\partial
V}{\partial q^j}) \mu^{a}_i + \dot{q}^i\dot{q}^j\frac{\partial
\mu^{a}_i}{\partial q^j}\right) Z^{b} 
\end{equation}
where $(C_{ab})$ is the inverse matrix of $(C^{ab})$ and 
the vector field $Z^{a}$ is defined by 
\[
g(Z^{a}, Y)=\mu^a(Y), \ \hbox{ for all vector field } Y, \; 1 \leq a \leq m\; ,
\]
that is, $Z^{a}$ is the gradient of the 1-form $\mu^{a}$.
Thus, ${\cal D}^{\perp}=\langle Z^{a}\rangle$, $1\leq a\leq m$.
In local coordinates, we have
\[
Z^{a} = g^{ij} \mu^{a}_i \frac{\partial}{\partial q^j} \; .
\]

By using the metric $g$ and the distribution ${\cal D}$ we can obtain
two complementary projectors  
\[
\begin{array}{rcl}
{\cal P}: TQ&\rightarrow&{\cal D}\; ,\\
{\cal Q}: TQ&\rightarrow&{\cal D}^{\perp}\; ,
\end{array}
\]
with respect to $g$.
The projector  ${\cal Q}$ is locally described by
\[
{\cal Q} = C_{ab} Z^a \otimes \mu^b \; .
\]

Using these projectors we can obtain the equations of motion as follows. A curve $c(t)$ is a motion for the non-holonomic system if
it satisfies the constraints, say, $\phi^{a}(\dot{c}(t))=0$, for all
$a$, and, in addition, the ``projected equation of motion''
\begin{equation}\label{q5}
{\cal P}(\nabla_{\dot{c}(t)} ~ \dot{c}(t)) =
- {\cal P}(\hbox{grad} ~ V(c(t)))
\end{equation}
is fulfilled. 
But these conditions are equivalent to
\[
\dot{c}(t) \in {\cal D}_{c(t)} \; , \; 
\bar{\nabla}_{\dot{c}(t)}\dot{c}(t) = - {\cal P}({\rm grad} ~ V(c(t)))\; ,
\]
where $\bar{\nabla}$ is the modified linear connection defined by
\[
\bar{\nabla}_X Y = \nabla_X Y + (\nabla_X {\cal Q})(Y) 
\]
for all vector fields $X$ and $Y$ on $Q$.

Since the constraints are linear then, from (\ref{qwe})
\begin{equation}\label{qwe1}
\hskip-3cm-\mu_i^a(q_0)g^{ij}(q_0)\frac{\partial S^h}{\partial q^j_0}(q_0, q_1)+\mu^a_i(q_0)g^{ij}(q_0)\int^h_0 \widetilde{\Lambda}(q(t),\dot{q}(t))\frac{\partial q}{\partial q^j_0}=0\; ,\quad 1\leq a\leq m\; ,
\end{equation}
or, in terms of projectors,
\begin{equation}\label{const111}
 {\cal Q}_{|q_0}\left( D_1 S^h(q_0, q_1)\right))={\cal Q}_{|q_0}\left( D_1 \int^h_0 \widetilde{\Lambda}(q(t),\dot{q}(t))\right)
\end{equation}
Moreover, the dynamics preserves the constraints $\Psi^a$ which implies that
\[
\Psi^a(q_1,   \frac{\partial S^h}{\partial q_1}(q_0, q_1)-\int^{h}_{0} \widetilde{\Lambda}(q(t),\dot{q}(t))\frac{\partial q}{\partial q_1})=0\; ,
\]
or, in other words, 
\begin{equation}\label{const222}
 {\cal Q}_{|q_1}\left( D_2 S^h(q_0, q_1)\right)={\cal Q}_{|q_1}\left( D_2 \int^h_0 \widetilde{\Lambda}(q(t),\dot{q}(t)\right)
\end{equation}
Therefore,  equations (\ref{const111}) and (\ref{const222}) show that the preservation of the exact constraints is equivalent 
to give a prescription about the relationship between the ``generating function" and the constraint forces. 

Thus, equations (\ref{ase}) 
\[
\hskip-2cm D_2 S^h(q_{k-1}, q_k)+D_1 S^h(q_k, q_{k+1})=
\int^h_0\widetilde{\Lambda}(q(t), \dot{q}(t))\frac{\partial q}{\partial q_1}+\int^{2h}_h\widetilde{\Lambda}(q(t), \dot{q}(t))\frac{\partial q}{\partial q_0}\; ,
\]
can be rewritten using expression (\ref{const222}) as follows
\begin{equation}\label{ase1}
{\cal P}_{|q_k}\left(D_2 S^h(q_{k-1}, q_k)\right)+D_1 S^h(q_k, q_{k+1})=
{\cal P}_{|q_k}\left(\int^h_0\widetilde{\Lambda}(q(t), \dot{q}(t))\frac{\partial q}{\partial q_1}\right)+\int^{2h}_h\widetilde{\Lambda}(q(t), \dot{q}(t))\frac{\partial q}{\partial q_0}\; ,
\end{equation}

Now, considering an approximated generating function $\tilde{S}^h$ and an approximate constraint force $\alpha^h=\alpha^h_0 (q_0,q_1)\, dq_0+\alpha^h_1(q_0, q_1)\, dq_1$, as in Remark \ref{wer}, from the previous discussion, we now substitute the approximated constraint force by: 
\begin{eqnarray*}
\tilde{\alpha}^h&=&\alpha^h_0 (q_0,q_1)\, dq_0\\
&&
+
{\cal P}_{|q_1}(
\alpha^h_1(q_0, q_1)\, dq_1)+ {\cal Q}_{|q_1}\left( D_2 \tilde{S}^h(q_0, q_1)\right))
\end{eqnarray*}

Therefore for $\tilde{S}^h$ and $\tilde{\alpha}^h$ equations (\ref{ase2}) are rewritten as
\begin{equation}\label{ase222}
{\cal P}_{|q_k}\left(D_2 \tilde{S}^h(q_{k-1}, q_k)\right)+D_1 \tilde{S}^h(q_k, q_{k+1})=
{\cal P}_{|q_k}\left( \alpha^h_1(q_{k-1}, q_k)\right)+\alpha^h_0(q_k, q_{k+1}),
\end{equation}
 for $1\leq k\leq N-1$. 
The importance of equations (\ref{ase222}) is that they generate an algorithm which automaticaly preserves the exact constraint functions $\Phi^a$.
In fact, if we apply the projector ${\cal Q}$ to Equations (\ref{ase222}) we obtain:
\begin{equation}\label{const1111}
 {\cal Q}_{|q_{k}}\left( D_1 S^h(q_k, q_{k+1})\right)={\cal Q}_{|q_k}\left( \alpha^h_0(q_k, q_{k+1}\right)
\end{equation}
or 
\[
\tilde{\varphi}^a(q_{k}, q_{k+1}, h)=\Psi^a(q_{k}, -\frac{\partial \tilde{S}^h}{\partial q_0}(q_k, q_{k+1})+\alpha^h_0(q_{k}, q_{k+1}))=0
\]
that is, the constraints are satisfied.

Therefore the geometric algorithm that we have obtained work as follows:

\fbox{\parbox{16cm}{
\[
\hskip-2cm{\cal P}_{|q_k}\left(D_2 \tilde{S}^h(q_{k-1}, q_k)\right)+D_1 \tilde{S}^h(q_k, q_{k+1})=
{\cal P}_{|q_k}\left( \alpha^h_1(q_{k-1}, q_k)\right)+\alpha^h_0(q_k, q_{k+1}),
\]
with initial condition satisfying: 
\[
\tilde{\varphi}^a(q_0, q_1, h)=0
\]
}}

Choosing  $\alpha^h_0$ and $\alpha^h_1$  in ${\cal D}^0$, we obtain equations for nonholonomic integrators with more geometric flavour:

\fbox{\parbox{16cm}{
{\bf Geometric  nonholonomic integrator}
\[
{\cal P}_{|q_k}\left(D_2 \tilde{S}^h(q_{k-1}, q_k)+D_1 \tilde{S}^h(q_k, q_{k+1})\right)=0
\]
}}

\noindent which is interpreted as a discretization of Equations (\ref{q5})
\[
\bar{\nabla}_{\dot{c}(t)}\dot{c}(t)=-{\cal P}(\hbox{grad}\,(V(c(t)))
\]
In a future work we will study from numerical and geometrical points of view this particular subclass of geometric integrators.

\subsection{Nonholonomic integrators preserving constraints}
For the class of integrators introduced in Section \ref{nh}, we find the following family of nonholonomic integrators preserving constraints: 
\begin{eqnarray*}\label{ase11}
{\cal P}_{|q_k}\left(D_2 S^h_{\alpha}(q_{k-1}, q_k)\right)+D_1 S^h_{\alpha}(q_k, q_{k+1})&=&
\alpha h {\cal P}_{|q_{k}}\left(\widetilde{\Lambda}((1-\alpha)q_{k-1}+\alpha q_k, \frac{q_k-q_{k-1}}{h})\right)\\
&&\hskip-2cm +(1-\alpha) h \widetilde{\Lambda}((1-\alpha)q_k+\alpha q_{k+1}, \frac{q_{k+1}-q_k}{h})\; ,\quad 1\leq k\leq N-1\; ,
\end{eqnarray*}
with initial condition satisfying 
\[
-\mu^a_i(q_0)g^{ij}(q_0)\frac{\partial S^h_{\alpha}}{\partial q^j_0}(q_0, q_1)+ 
(1-\alpha) h \mu^a_i(q_0)g^{ij}(q_0)\widetilde{\Lambda}_j((1-\alpha)q_0+\alpha q_1, \frac{q_1-q_0}{h}))=0\; .
\]

\begin{example}[The nonholonomic particle revisited]
 {\rm 

\begin{eqnarray*}
&&\hskip-0.5cm\frac{1}{1+y_1^2}\left(\frac{x_1-x_0}{h}-h\frac{x_1+x_0}{2}\right)-\frac{x_2-x_1}{h}-h\frac{x_2+x_1}{2}+\frac{y_1}{1+y_1^2}\left(\frac{z_1-z_0}{h}\right)\\
&&\hskip-0.3cm= -\frac{h}{2}\left[
\frac{
\frac{1}{1+y_1^2}\cdot\frac{(x_1+x_0)(y_1+y_0)}{2}
-\frac{(x_1-x_0)(y_1-y_0)}{h^2}}
{1+\left(\frac{y_1+y_0}{2}\right)^2}\cdot \frac{y_1+y_0}{2}
+\frac{
\frac{(x_2+x_1)(y_2+y_1)}{2}-\frac{(x_2-x_1)(y_2-y_1)}{h^2}
}
{1+\left(\frac{y_2+y_1}{2}\right)^2}\cdot \frac{y_2+y_1}{2}
\right.\\
&&\left.-\frac{y_1}{1+y_1^2}\frac{
\frac{(x_1+x_0)(y_1+y_0)}{2}-\frac{(x_1-x_0)(y_1-y_0)}{h^2}
}
{1+\left(\frac{y_1+y_0}{2}\right)^2}
\right]\\
&&\hskip-0.5cm\frac{y_1-y_0}{h}-h\frac{y_1+y_0}{2}-\frac{y_2-y_1}{h}-h\frac{y_2+y_1}{2}=0\; ,\\
&&\hskip-0.5cm\frac{y_1^2}{1+y_1^2}\left(\frac{z_1-z_0}{h}\right)-\frac{z_2-z_1}{h}+\frac{y_1}{1+y_1^2}\left( 
\frac{x_1-x_0}{h}-h\frac{x_1+x_0}{2}\right)
\\
&&\hskip-0.3cm= \frac{h}{2}\left[
\frac{y_1^2}{1+y_1^2}\frac{
\frac{(x_1+x_0)(y_1+y_0)}{2}
-\frac{(x_1-x_0)(y_1-y_0)}{h^2}}
{1+\left(\frac{y_1+y_0}{2}\right)^2}
+\frac{
\frac{(x_2+x_1)(y_2+y_1)}{2}-\frac{(x_2-x_1)(y_2-y_1)}{h^2}
}
{1+\left(\frac{y_2+y_1}{2}\right)^2}\right.\\
&&\left. -\frac{y_1}{1+y_1^2}\frac{
\frac{(x_1+x_0)(y_1+y_0)}{2}-\frac{(x_1-x_0)(y_1-y_0)}{h^2}
}
{1+\left(\frac{y_1+y_0}{2}\right)^2}\cdot \frac{y_1+y_0}{2}\right]\; .\\
\end{eqnarray*}
with initial condition satisfying
\begin{eqnarray*}
&&\hskip-0.5cm\tilde{\varphi}^a(x_0, y_0, z_0, x_1, y_1, z_1, h)=
-\frac{z_1-z_0}{h}-\frac{h}{2}
\frac{
\frac{(x_1+x_0)(y_1+y_0)}{2}
-\frac{(x_1-x_0)(y_1-y_0)}{h^2}}
{1+\left(\frac{y_1+y_0}{2}\right)^2}
\\&+&y_0\left[
\frac{x_1-x_0}{h}+h\frac{x_1+x_0}{2}
-
\frac{h}{2}
\frac{
\frac{(x_1+x_0)(y_1+y_0)}{2}
-\frac{(x_1-x_0)(y_1-y_0)}{h^2}}
{1+\left(\frac{y_1+y_0}{2}\right)^2}\cdot \frac{y_1+y_0}{2}
\right]\; .
\end{eqnarray*}
For the same initial conditions and data, the following graph shows the exact preservation   of the constraint function evolution with time of our algorithm.

\begin{center}
\includegraphics[width=10cm]{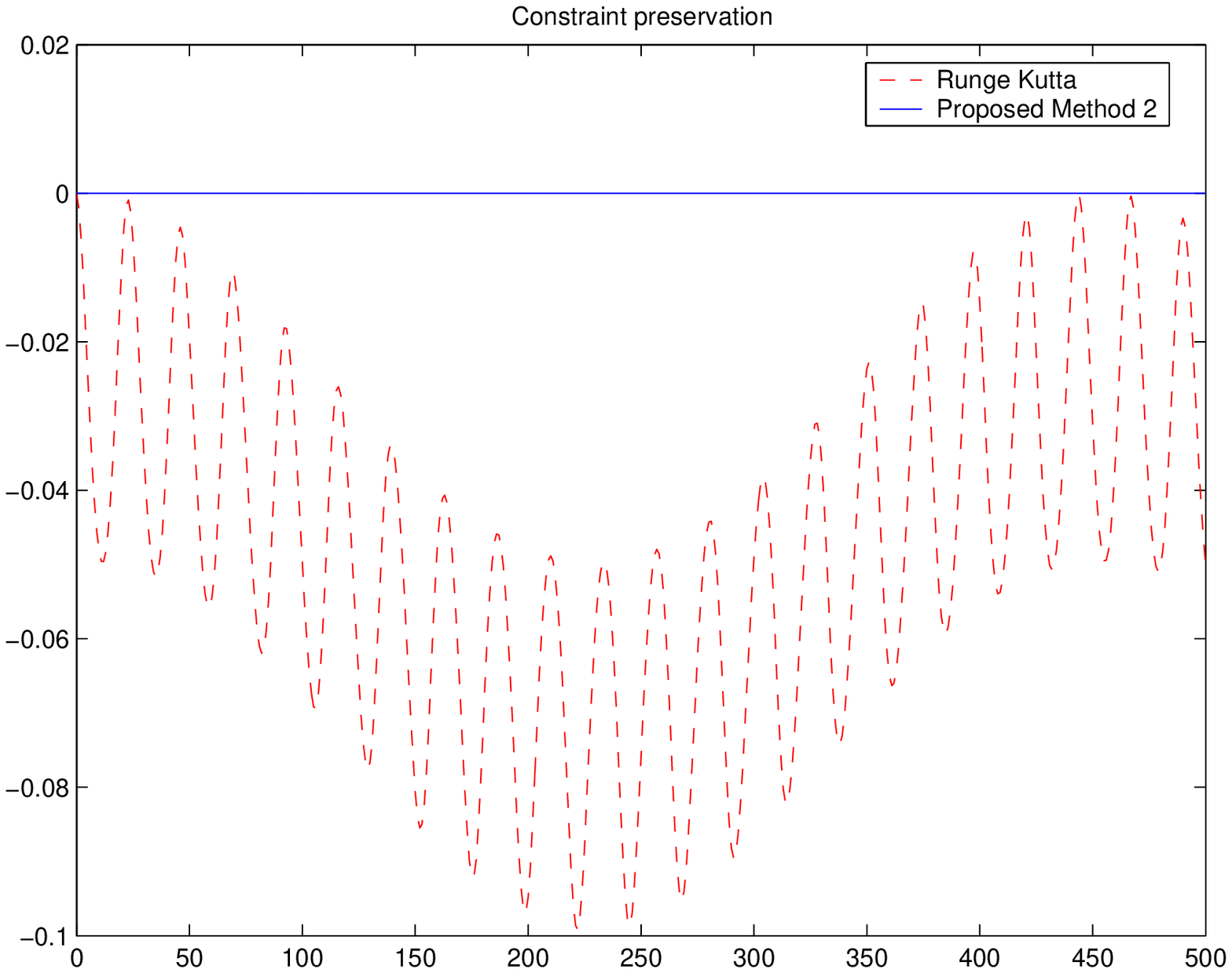}
\end{center}

}
\end{example}

\section{Conclusion}

A new numerical algorithm has been  proposed for nonholonomic mechanics. This algorithm is based in the underlying geometry of nonholonomic systems. For mechanical systems with linear constraints, a geometric integrator preserving constraints is proposed. 
 
In  future work, we will explore reduction schemes for discrete systems using the approach of generating functions. It is also interesting to use generating functions of different kinds; in a recent work \cite{LDA1}, we have shown that generating functions of second class generate   algorithms   which are symplectic (in some sense) for discrete optimal control theory (see also \cite{LDA}).  Moreover, we may easily extend the generating function technique in order to consider variable time stepping and also the time-dependent case and it would be possible to use this formalism for classical field theories.

\section*{Acknowledgments}

This work has been  supported by grant BFM2001-2272. A. Santamar{\'\i}a Merino wishes to thank the Programa de formaci\'on de Investigadores of the Departamento de Educaci\'on, Universidades e Investigaci\'on of the Basque Government (Spain) for financial support.

\end{document}